\documentclass[conference]{IEEEtran}
\usepackage{cite}
\usepackage{amsmath,amssymb,amsfonts}
\usepackage{graphicx}
\usepackage{textcomp}
\usepackage{xcolor}
\def\BibTeX{{\rm B\kern-.05em{\sc i\kern-.025em b}\kern-.08em
    T\kern-.1667em\lower.7ex\hbox{E}\kern-.125emX}}

\usepackage{ifthen}
\usepackage{tikz}
\usepackage{pgfplots,pgfplotstable}
\pgfplotsset{
        set layers={
            background,
            main,
        },
    }
\newlength{\figheight}
\usepackage[square,sort,comma,numbers]{natbib}
\usepackage[T1]{fontenc}

\usetikzlibrary{calc, patterns, spy}

\usepackage{listings}
\AtBeginDocument{\DeclareCaptionSubType{lstlisting}}
\usepackage{caption,subcaption}
\usepackage[binary-units=true]{siunitx} 
\usepackage{booktabs}
\usepackage[colorlinks=true,linkcolor=black,anchorcolor=black,citecolor=black,filecolor=black,menucolor=black,runcolor=black,urlcolor=black]{hyperref}
\setcitestyle{square}
\usepackage{multirow}

\usepackage[ruled, lined, linesnumbered, commentsnumbered, longend]{algorithm2e}

\usepackage[acronym,nonumberlist]{glossaries}
\newacronym[longplural={Deep Neural Networks}]{DNN}{DNN}{Deep Neural Network}
\newacronym{DL}{DL}{Deep Learning}
\newacronym{OSL}{OSL}{One-Shot Learning}
\newacronym{BNN}{BNN}{Binary Neural Network}
\newacronym{CIM}{CIM}{computing-in-memory}
\newacronym[longplural={Content-addressable memories}]{CAM}{CAM}{Content-addressable memory}
\newacronym[longplural={non-volatile memories}]{NVM}{NVM}{non-volatile memory}
\newacronym{ReRAM}{ReRAM}{Resistive RAM}
\newacronym{MRAM}{MRAM}{Magnetoresistive RAM}
\newacronym{FeFET}{FeFET}{Ferroelectric FET}
\newacronym{ACAM}{ACAM}{analog CAM} 
\newacronym{MCAM}{MCAM}{multi-state CAM}
\newacronym{HD}{HD}{Hamming Distance}
\newacronym{MCA}{MCA}{Memristive Crossbar Array}
\newacronym{DAC}{DAC}{digital-to-analog converter}
\newacronym{ADC}{ADC}{analog-to-digital converter}
\newacronym{AP}{AP}{Associative Processor}
\newacronym{RTM}{RTM}{Racetrack Memory}
\newacronym{CSE}{CSE}{Common Subexpression Elimination}
\newacronym{DFG}{DFG}{Data-flow graph}
\newacronym{MVM}{MVM}{matrix-vector multiplication}

\usepackage{svg}

\usepackage{amsmath}
\usepackage{nicematrix}

\definecolor{pairedOneLightBlue}{HTML}{a6cee3}
\definecolor{pairedTwoDarkBlue}{HTML}{1f78b4}
\definecolor{pairedThreeLightGreen}{HTML}{b2df8a}
\definecolor{pairedFourDarkGreen}{HTML}{33a02c}
\definecolor{pairedFiveLightRed}{HTML}{fb9a99}
\definecolor{pairedSixDarkRed}{HTML}{e31a1c}
\definecolor{butter1}{rgb}{0.988,0.914,0.310}
\definecolor{butter2}{rgb}{0.929,0.831,0.000}
\definecolor{butter3}{rgb}{0.769,0.627,0.000}
\definecolor{orange1}{rgb}{0.988,0.686,0.243}
\definecolor{orange2}{rgb}{0.961,0.475,0.000}
\definecolor{orange3}{rgb}{0.808,0.361,0.000}
\definecolor{chocolate1}{rgb}{0.914,0.725,0.431}
\definecolor{chocolate2}{rgb}{0.757,0.490,0.067}
\definecolor{chocolate3}{rgb}{0.561,0.349,0.008}
\definecolor{chameleon1}{rgb}{0.541,0.886,0.204}
\definecolor{chameleon2}{rgb}{0.451,0.824,0.086}
\definecolor{chameleon3}{rgb}{0.306,0.604,0.024}
\definecolor{skyblue1}{rgb}{0.447,0.624,0.812}
\definecolor{skyblue2}{rgb}{0.204,0.396,0.643}
\definecolor{skyblue3}{rgb}{0.125,0.290,0.529}
\definecolor{plum1}{rgb}{0.678,0.498,0.659}
\definecolor{plum2}{rgb}{0.459,0.314,0.482}
\definecolor{plum3}{rgb}{0.361,0.208,0.400}
\definecolor{scarletred1}{rgb}{0.937,0.161,0.161}
\definecolor{scarletred2}{rgb}{0.800,0.000,0.000}
\definecolor{scarletred3}{rgb}{0.643,0.000,0.000}
\definecolor{aluminium1}{rgb}{0.933,0.933,0.925}
\definecolor{aluminium2}{rgb}{0.827,0.843,0.812}
\definecolor{aluminium3}{rgb}{0.729,0.741,0.714}
\definecolor{aluminium4}{rgb}{0.533,0.541,0.522}
\definecolor{aluminium5}{rgb}{0.333,0.341,0.325}
\definecolor{aluminium6}{rgb}{0.180,0.204,0.212}

\definecolor{blind_safe_one_scheme_three_colors}{RGB}{102,194,165}
\definecolor{blind_safe_two_scheme_three_colors}{RGB}{252,141,98}
\definecolor{blind_safe_three_scheme_three_colors}{RGB}{141,160,203}

\definecolor{blind_safe_one_scheme_four_colors}{RGB}{166,206,227}
\definecolor{blind_safe_two_scheme_four_colors}{RGB}{31,120,180}
\definecolor{blind_safe_three_scheme_four_colors}{RGB}{178,223,138}
\definecolor{blind_safe_four_scheme_four_colors}{RGB}{51,160,44}

\definecolor{blind_safe_one_scheme_five_colors}{RGB}{240,249,232}
\definecolor{blind_safe_two_scheme_five_colors}{RGB}{186,228,188}
\definecolor{blind_safe_three_scheme_five_colors}{RGB}{123,204,196}
\definecolor{blind_safe_four_scheme_five_colors}{RGB}{67,162,202}
\definecolor{blind_safe_five_scheme_five_colors}{RGB}{8,104,172}

\definecolor{blind_safe_one_scheme_seven_colors}{RGB}{118,42,131}
\definecolor{blind_safe_two_scheme_seven_colors}{RGB}{175,141,195}
\definecolor{blind_safe_three_scheme_seven_colors}{RGB}{231,212,232}
\definecolor{blind_safe_four_scheme_seven_colors}{RGB}{247,247,247}
\definecolor{blind_safe_five_scheme_seven_colors}{RGB}{217,240,211}
\definecolor{blind_safe_six_scheme_seven_colors}{RGB}{127,191,123}
\definecolor{blind_safe_seven_scheme_seven_colors}{RGB}{27,120,55}
\definecolor{blind_safe_eight_scheme_seven_colors}{RGB}{66,102,245}

\definecolor{yellow_one}{RGB}{255,255,212}
\definecolor{yellow_two}{RGB}{254,217,142}
\definecolor{yellow_three}{RGB}{254,153,41}
\definecolor{yellow_four}{RGB}{217,95,14}
\definecolor{yellow_five}{RGB}{153,52,4}

\definecolor{highlight_code_color1}{RGB}{199,233,180}
\definecolor{highlight_code_color2}{RGB}{254,196,79}

\begin{document}
\title{Full-Stack Optimization for CAM-Only DNN Inference}

\author{\IEEEauthorblockN{João Paulo C. de Lima\textsuperscript{*$\dagger$$\ddagger$}, Asif Ali Khan\textsuperscript{*}, Luigi Carro\textsuperscript{$\ddagger$}, Jeronimo Castrillon\textsuperscript{*$\dagger$}}
\IEEEauthorblockA{\textsuperscript{*}Chair for Compiler Construction, Technische Universität Dresden, Dresden, Germany\\
 \textsuperscript{$\dagger$}Center for Scalable Data Analytics and Artificial Intelligence (ScaDS.AI), Dresden, Germany \\ 
  \textsuperscript{$\ddagger$}Informatics Institute, Federal University of Rio Grande do Sul, Porto Alegre, Brazil  \\
Email: \{joao.lima, asif\_ali.khan, jeronimo.castrillon\}@tu-dresden.de, carro@inf.ufrgs.br}
}


\maketitle

\begin{abstract}
The accuracy of neural networks has greatly improved across various domains over the past years. 
Their ever-increasing complexity, however, leads to prohibitively high energy demands and latency in von-Neumann systems. 
Several \emph{computing-in-memory} (CIM) systems have recently been proposed to overcome this, but trade-offs involving accuracy, hardware reliability, and scalability for large models remain a challenge.
Additionally, for some CIM designs, the activation movement still requires considerable time and energy.   
This paper explores the combination of algorithmic optimizations for ternary weight neural networks and \emph{associative processors} (APs) implemented using \emph{racetrack memory} (RTM).
We propose a novel compilation flow to optimize convolutions on APs by reducing their arithmetic intensity. 
By leveraging the benefits of RTM-based APs, this approach substantially reduces data transfers within the memory while addressing accuracy, energy efficiency, and reliability concerns. 
Concretely, our solution improves the energy efficiency of ResNet-18 inference on ImageNet by 7.5$\times$ compared to crossbar in-memory accelerators while retaining software accuracy.


\end{abstract}

\section{Introduction}

Neural network (NN) models have experienced growth, leading to enhanced accuracy and wider applicability across diverse domains. However, this progress has also brought forth computational challenges due to the substantial data movement between memory and processing units. 
To mitigate this, algorithmic approaches for efficient NN acceleration, such as pruning, extreme quantization, and optimized convolution algorithms like Winograd, can reduce memory requirements and computational costs, while, in the architectural front, the \emph{computing-in-memory} (CIM) paradigm exploits the physical attributes of the memory cells to compute in place~\cite{smagulova2023resistive}.  

Matrix multiplication is a crucial operation in NNs, and CIM-based accelerators often use resistive crossbar arrays for analog/mixed-signal matrix-vector multiplication (MVM) in constant time, achieving significant performance gains over digital CMOS methods. 
However, their widespread usage is hindered by reliability issues and energy-hungry digital-to-analog (DAC) and analog-to-digital (ADC) conversions. 
Despite numerous advances in ADCs/DACs~\cite{smagulova2023resistive}, crossbars still exhibit high inaccuracy, are limited to small-scale use cases, and face reliability issues. 
As alternative to crossbars, promising CIM paradigms like \emph{content-addressable memories} (CAMs) have also been explored \cite{choi2018content, ran2023pecan, nguyen2023deepcam}. 
Despite their excellent performance and no need for costly peripherals, current CAM designs struggle to maintain accuracy for larger networks as they rely on methods like binarized networks (BNNs)~\cite{choi2018content}, approximate MACs~\cite{imani2017efficient}, or dot-product approximations~\cite{ran2023pecan, nguyen2023deepcam}.

Recent research has shown that ternary weights networks (TWNs), i.e., $w_i \in \{-1,0,1\}$, combined with reduced-precision activations (often integers with 8 to 4 bits) can retain accuracy while drastically reducing computational complexity on specialized hardware~\cite{esser2019learned, diffenderfer2021multi}. In bulk-bitwise CIM designs, if executed intelligently, such ternary models would rely solely on $O(m)$ addition/subtraction, eliminating the need for $O(m^{2})$ multiplication, where $m$ is the bitwidth of operands. 
However, existing CIM designs prioritize efficient handling of weights, disregarding the significant cost of moving activations.
The cost of transferring activations can become significant, depending on the activation precision, possibly outweighing the cost of weights.
In modern NNs, this movement can consume 30\% to 70\% of the total CIM system energy~\cite{scherer2021cutie}. 
 
In this paper, we present a full-stack solution, comprising a compilation flow and CAM-based accelerator, that considers accuracy as a first-class optimization metric, along with performance and energy consumption.  
By consolidating previously scattered methods, we investigate their joint impact and propose a sequence of steps to co-optimize bulk-bitwise convolutions for performance and to reduce activation transfer.
Specifically, we employ \emph{racetrack memory} (RTM)-based CAMs to implement \emph{associative processors} (APs) for DNN inference. Compared to other NVMs, RTMs~\cite{blasing2020magnetic} provide an order of magnitude higher density, comparable performance, and improved energy efficiency and endurance.
Concretely, we make the following novel contributions: 
\begin{itemize}
    \item An RTM-based AP for efficient bulk-bitwise processing. By sequentially storing feature maps in RTM cells, we effectively leverage the sequential accesses of RTMs for bit-serial word-parallel convolution in the AP model. 
    \item A compilation flow for the proposed AP with optimizations to reduce the arithmetic intensity of convolution and transform weights into AP instructions. 
    Our transformation focuses on eliminating multiplications, removing redundant additions/subtractions and optimizing the bitwidth of partial sums. 
   \item We evaluate our full-stack solution on various NN architectures - VGG9, VGG11, and ResNet18, using CIFAR10 and ImageNet datasets. Ensuring software accuracy, our RTM-APs accelerate inference for the largest model (ResNet-18/ImageNet) by 3$\times$ with 2.5$\times$ lower energy consumption compared to crossbar-based accelerators, resulting in 7.5$\times$ energy efficiency improvement.
   
\end{itemize}

\section{Background}
\label{sec:background}
This section provides a concise overview of the RTM technology, CAMs and APs, as well as neural networks.

\subsection{DNNs: Quantization and sparsity}
\label{subsec:btnns}
\begin{figure}[tbh]
\centering
\vspace{-0.3cm}
\includegraphics[scale=0.5]{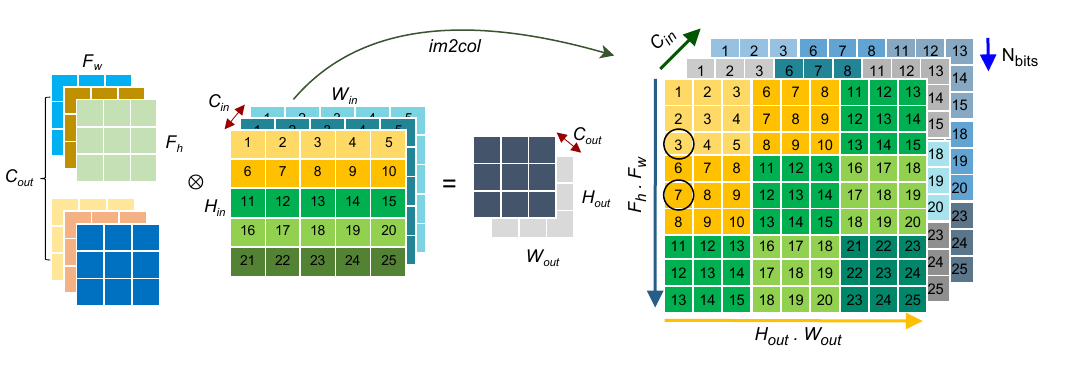}
\vspace{-0.5cm}
\caption{Direct convolution and im2col transformation}
\label{fig_nn}
\vspace{-0.3cm}
\end{figure}

Convolutional layers, depicted in Fig.~\ref{fig_nn}, are the most computationally intensive in modern DNNs. They take multiple $C_{in}$ \textit{input feature maps} (IFMs) and convolve them with $C_{out}$ \textit{filters} to generate $C_{out}$ \textit{output feature maps} (OFMs).

Most DNNs are overparameterized and contain several sub-networks with comparable accuracy and substantially less memory and computational intensity.
This concept backs popular methods like pruning and quantization.
The recently proposed multi-prize lottery ticket hypothesis~\cite{diffenderfer2021multi} also supports the existence of an optimal binary sub-network inside a super-network. 
This finding establishes a necessary condition for accurately implementing NN hardware using low-cost bitwise operations, like the XNOR-popcount implementation of BNNs~\cite{choi2018content}. 
However, the activation requirements tell a different story: binary activations strongly limit the representational capacity of the network, even when taking batch normalization into account~\cite{diffenderfer2021multi}. 
As a consequence, this results in an unavoidable loss in accuracy. Recent research shows that a moderate activation quantization, such as 4-bit, is sufficient to maintain the same accuracy as its full-precision counterpart, even in complex classification problems~\cite{esser2019learned}.

\subsection{Content addressable memories and associative processing}
\label{subsec:cams}
CAMs represent a class of CIM that enables fast and energy-efficient memory search operations. 
CAMs take an input pattern (search key), compare it with all the contents of the CAM array, and return locations where the contents match. 
For the search operation, all match lines are precharged to a high voltage and the key is applied to the CAM input lines. 
If a single cell mismatch occurs, the corresponding match line discharges. Otherwise, the match line will stay high when all bits are matched.
The match lines are connected to a sensing circuitry to decode a corresponding address.  
This way, parallel searches are performed on all CAM content in $O(1)$. 

\noindent{\textbf{Associative processors (APs):}} In various studies, CAMs have been used for exact and approximate search~\cite{nguyen2023deepcam,ran2023pecan}, as well as for bulk-bitwise arithmetic and logic operations, also known as APs \cite{zha2020hyper}.
APs perform in-place computation on stored inputs using CAM search/write, specifically masked search and parallel column write operations.
In this execution model, the truth table of a Boolean expression is broken down into simpler search and write primitives, generating a lookup table (LUT) of masked keys to implement the required function. 
Fig.~\ref{fig_mechanism}c illustrates an AP consisting of a CAM array, which allows searching a key in the stored content and writing a data pattern (write key) in all tagged rows (see Sec.~\ref{sec:rtm-ap}).
Depending on the operation (search/write), the AP controller sets the mask and key values iteratively by referencing the corresponding LUT (Table~\ref{fig_lut}). 
In the \textit{search phase}, the search key and mask ﬁelds are set according to the left side of the LUT and compared with the CAM content in the speciﬁed/masked columns and all rows in parallel.  
The matched lines are stored in the \textit{tag register}, which drives updating the data pattern in all tagged rows in the next phase. 
In the \textit{write phase}, the mask and key values are set by observing the LUT's right side, and the CAM content in the tagged rows and selected columns is written accordingly. 
Thanks to the parallel search across all rows, any function performed on a sequential processor can be implemented on the AP in a SIMD fashion.

\subsection{Racetrack memory}
\label{subsec:rtm}
RTM is a magnetic NVM technology where a single cell comprises a magnetic nanowire, also referred to as a track, with the capacity to store up to 100 data bits (domains) (see Fig.~\ref{fig_mechanism}e)~\cite{junsangsri2016non}. Each track has one or more access ports that enable the read/write operations. The domain walls must be first shifted and aligned with the access port to access a specific domain within a track. Typically, multiple tracks are grouped into domain-wall block clusters (DBCs) and can be accessed in parallel. 
While RTMs offer higher storage density and lower power operations than other memory technologies~\cite{blasing2020magnetic},  their read/write speed can vary with different access patterns due to the movement of magnetic domain walls in nanowires.
For this reason, data in RTM is typically stored in a bit-interleaved fashion to reduce the number of shifts and sequential accesses.

\section{RTM-AP: Accelerator architecture}
\label{sec:rtm-ap}
\begin{figure*}[h]
\centering
\includegraphics[scale=0.7]{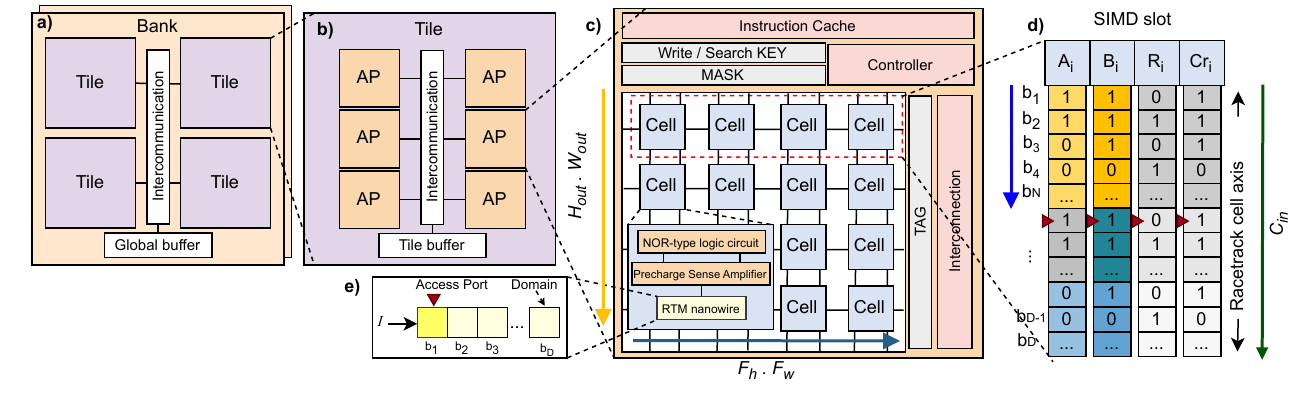}
\vspace{-0.4cm}
\caption{RTM-AP. a) Hierarchical accelerator architecture consisting of banks, tiles, APs, buffers and interconnection network, b) tile showing array of APs, c) AP organization consisting of CAM array, registers, instruction cache and control unit, d) SIMD slot for two-operand addition and mapping of inputs to racetrack domains, e) an RTM nanowire.} 
\label{fig_mechanism}
\vspace{-0.5cm}
\end{figure*}

As illustrated in Fig.~\ref{fig_mechanism}a-c, the accelerator is organized into a three-level hierarchy comprising banks, tiles, and APs. 
Each AP is an independent processing unit operated and accessed in parallel.
The hierarchical organization enables fine-grained control over the compute resources by allocating banks, tiles, and APs based on the computational needs of each layer.

\begin{table}
\fontsize{5}{6}\selectfont 
\setlength{\tabcolsep}{3pt} 
\centering
\caption{LUT tables for both in-place and out-of-place 1-bit addition (left) and subtraction (right).} \label{fig_lut}
\begin{tabular}{|p{0.0cm}p{0.0cm}p{0.0cm}|p{0.0cm}p{0.0cm}p{0.5cm}|p{0.0cm}p{0.0cm}p{0.5cm}|p{0.0cm}p{0.0cm}p{0.0cm}|p{0.0cm}p{0.0cm}p{0.5cm}|p{0.0cm}p{0.0cm}p{0.5cm}|}
\hline
\multicolumn{3}{|c}{\textbf{1b Adder}} & \multicolumn{3}{|c}{\textbf{In-place}} & \multicolumn{3}{|c}{\textbf{Out-of-place}} & \multicolumn{3}{|c}{\textbf{1b Sub}} & \multicolumn{3}{|c}{\textbf{In-place}} & \multicolumn{3}{|c|}{\textbf{Out-of-place}} \\
\multicolumn{1}{|c}{Cr} & \multicolumn{1}{c}{B} & \multicolumn{1}{c}{A} & \multicolumn{1}{|c}{Cr} & \multicolumn{1}{c}{B} & \multicolumn{1}{c}{Comment} & \multicolumn{1}{|c}{Cr} & \multicolumn{1}{c}{R} & \multicolumn{1}{c}{Comment} & \multicolumn{1}{|c}{Br} & \multicolumn{1}{c}{B} & \multicolumn{1}{c}{A} & \multicolumn{1}{|c}{Br} & \multicolumn{1}{c}{B} & \multicolumn{1}{c}{Comment} & \multicolumn{1}{|c}{Br} & \multicolumn{1}{c}{R} & \multicolumn{1}{c|}{Comment} \\
\hline
0 & 0 & 0 & 0 & 0 & NC & 0 & 0 & NC & 0 & 0 & 0 & 0 & 0 & NC & 0 & 0 & NC \\
0 & 0 & 1 & 0 & 1 & 2nd & 0 & 1 & 1st & 0 & 0 & 1 & 1 & 1 & 1st & 1 & 1 & 1st \\
0 & 1 & 0 & 0 & 1 & NC & 0 & 1 & 2nd & 0 & 1 & 0 & 0 & 1 & NC & 0 & 1 & 2nd \\
0 & 1 & 1 & 1 & 0 & 1st & 1 & 0 & NC & 0 & 1 & 1 & 0 & 0 & 2nd & 0 & 0 & NC \\
1 & 0 & 0 & 0 & 1 & 3rd & 0 & 1 & 3rd & 1 & 0 & 0 & 1 & 1 & 4th & 1 & 1 & 3rd \\
1 & 0 & 1 & 1 & 0 & NC & 1 & 0 & NC & 1 & 0 & 1 & 1 & 0 & NC & 1 & 0 & NC \\
1 & 1 & 0 & 1 & 0 & 4th & 1 & 0 & 4th & 1 & 1 & 0 & 0 & 0 & 3rd & 0 & 0 & 4th \\
1 & 1 & 1 & 1 & 1 & NC & 1 & 1 & 5th & 1 & 1 & 1 & 1 & 1 & NC & 1 & 1 & 5th \\
\hline
\end{tabular}
\vspace{-0.5cm}
\end{table}

The bitwise implementation of addition or subtraction can be in-place or out-of-place, meaning that the result is written into one of the inputs or a new location. 
Table~\ref{fig_lut} illustrates the LUTs for both versions.  
The run order indicated in the comment column in Table~\ref{fig_lut} must be followed to ensure the correctness of bitwise computations. 
Note that keys marked as NC (no change) do not alter any content in the CAM, hence no search/write is needed. 
In all tables, $R$ and $C_{r}/B_{r}$ represent the result and carry/borrow, respectively, and A and B are the inputs.
At a coarser level, the AP instruction cache drives the execution of application kernels by providing the sequence of instructions, operand locations, and their bitwidth.

Fig.~\ref{fig_mechanism}c shows the CAM cell incorporating a NOR-type logic circuit, a precharge sense amplifier, and a memory device (in this case, an RTM nanowire).
APs support only bitwise operations and require storing the input so that each row stores bits from different operands at the same bit position.
In prior research \cite{zha2020hyper}, each row also stores all operands' bits across multiple columns due to limitations in their CAM design.
In the case of NVM technologies like RRAM, where the resistance window is divided into ranges to represent different bits, their multi-level cell operation is incompatible with bitwise operations. 
Consequently, they must operate in a single bit per cell mode, significantly losing capacity.

However, RTMs can effectively provide the mechanism needed for AP implementation.
Earlier research has successfully implemented ultra-dense, low-power CAMs using RTM and skyrmion devices~\cite{junsangsri2016non,gnawali2018low,zhang2023sky} based on manipulation and movement of domain walls. 
Since operands in APs are processed in a bit-serial fashion, in this work, we store the operands sequentially in the nanowires.
In the example in Fig.~\ref{fig_mechanism}d, each column $(A,B,R,C_r)$ represents an operand, and the row value ``1101'' is the readout bits given by the state of the current domains aligned to access ports.
Unlike prior research that employed interleaving to work around RTMs' sequential nature~\cite{blasing2020magnetic}, our novel execution model harnesses this characteristic for true multi-bit storage and bitwise processing.

\section{Compilation framework for RTM-APs}
\label{sec:comp}

\begin{figure*}[h]
\centering
\includegraphics[scale=0.75]{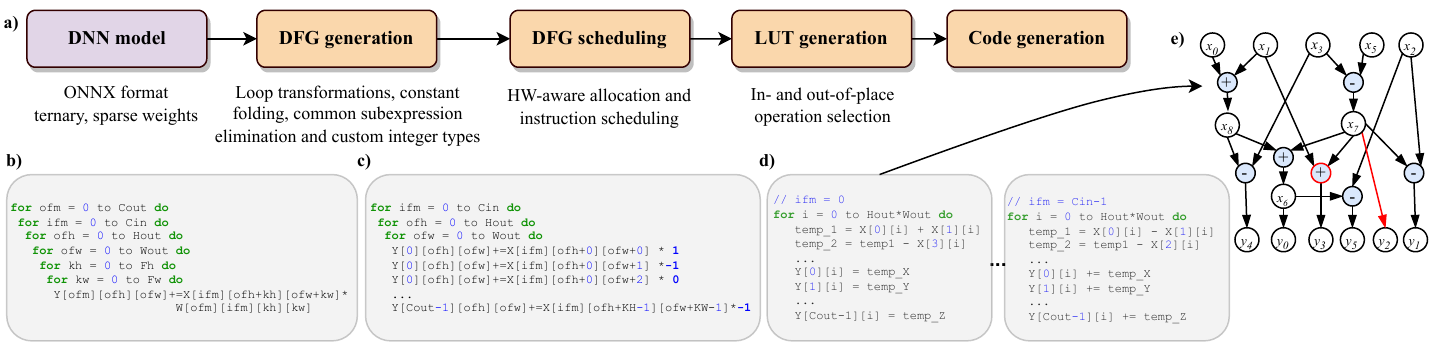}
\caption{a) Compilation flow and optimization techniques used in each step, b) naïve loop in convolutional layers, c) loop after applying loop interchange, unrolling and constant folding of ternary weights, d) loop after loop fission and common subexpression elimination, e) optimized data-flow graph (DFG) for Equation~\ref{eq:1}.} 
\label{fig_compilation}
\vspace{-0.6cm}
\end{figure*}

In bulk bitwise CIM architectures, we need to prioritize simpler and fewer operations, while other aspects important for von-Neumann systems, such as code size and temporal locality, may have less significance.
We propose an automatic compilation flow that optimizes convolutions at the application and arithmetic levels. 
This is achieved by automatically reducing the arithmetic intensity of operations and by efficiently mapping them to reduce data movement.
Fig.~\ref{fig_compilation}a depicts a high-level overview of our framework that transforms trained TWNs into an optimized sequence of AP instructions.   
In contrast to conventional convolution, where weights are fetched from memory and multiplied with inputs, our approach involves statically compiling weights into AP instructions. Additionally, our implementation stands apart from other CIM approaches that pass feature maps through multiple memory arrays, incurring high costs for peripheral circuits. Instead, we adopt a data-centric approach, where feature maps are primarily computed in place, significantly reducing data movement.

\subsection{Data-flow graph generation}

Consider the standard loop nest of convolutional layers shown in Fig.~\ref{fig_compilation}b. In inference tasks, $W$ can usually be known and fixed at compile time. Assuming trained TWNs, we can statically replace multiplications with expressions involving only additions and subtractions.
Depending on the model's sparsity, this can also eliminate many expressions involving multiplication with zero. 
To enable this optimization, we unroll loops $kh$ and $kw$ and apply constant weight folding, where weights limited to $\{-1, 0, 1\}$ replace multiplications. This optimization effectively reduces computational complexity by eliminating multiplication from the convolution kernel. However, it comes at the cost of fully unrolling all loops, leading to a substantial increase in code size overhead. 

To further reduce arithmetic complexity and code size, we must expose the loop body with the highest redundancy in additions before unrolling it. 
This region of high redundancy refers to the weight slices convolved on the same input patch. 
This is achieved by applying loop interchange on the naïve loop (Fig.~\ref{fig_compilation}b) to make $ofm$ the third innermost loop, followed by unrolling the three innermost loops—namely, the variables $ofm$, $kh$, and $kw$. After applying constant weight folding, the resultant loop has a larger body but a higher number of subexpressions depending only on an input patch of size $F_{h} \times F_{w}$, as illustrated in Fig.~\ref{fig_compilation}c.
Loop interchange is an enabler for a later optimization step, as using it alone is not beneficial due to poor locality. 
Initially, all three steps might appear counterintuitive, especially considering that real-world networks often have $C_{out}$ ranging from 64 to 512 and their overhead in code size and poor locality in CPUs.
However, these transformations target in-memory architectures, which differ significantly from von-Neumann ones, and existing compilers would not naturally explore these optimization steps.

The outermost loop from Fig.~\ref{fig_compilation}c is also fully unrolled, resulting in $C_{in}$ loop bodies depicted in Fig.~\ref{fig_compilation}d. Each of these bodies handles a single IFM and iterates over $ofh$ and $ofw$ variables, which are then simplified in a single loop over $H_{out}*W_{out}$.
They can be further optimized by \gls{CSE}, reducing the number of additions. 
\gls{CSE} aims to identify and eliminate common expressions within a program, reducing the total number of computations required and improving the program's overall performance. 
In this approach, \glspl{CSE} are found within the weight slice ($C_{out} \times 1 \times F_{h} \times F_{w}$), which is the slice that allows the greatest potential for reuse of a single input-channel across all output channels.
Equation~\ref{eq:1} illustrates a simple implementation of MVM with ternary weights. 
\begin{gather} \label{eq:1}
\tiny
 y
 =
  \begin{pNiceMatrix}[r]
   1 & -1 & 0 & 1 & 0 & -1 \\
   0 & 0 & -1 & 1 & \phantom{-}0 & -1 \\
   0 & 0 & 0 & -1 & 0 & 1 \\
   0 & -1 & 0 & -1 & 0 & 1 \\
   1 & -1 & 0 & -1 & 0 & 0 \\
   1 & -1 & -1 & 1 & 0 & -1 \\
   \end{pNiceMatrix}
  *
  \begin{pmatrix}
    x_{0} \\
    x_{1} \\
    x_{2} \\
    x_{3} \\
    x_{4} \\
    x_{5} \\
   \end{pmatrix}
   =
   \begin{pmatrix}
    x_{6} \\
    x_{7}-x_{2} \\
    -x_{7} \\
    -x_{1}-x_{7} \\  
    x_{8}-x_{3} \\
    x_{6}-x_{2} \\
   \end{pmatrix}  
\end{gather}

with $x_{6} = x_{7}+x_{8}$, $x_{7} = x_{3}-x_{5}$, $x_{8} = x_{0} + x_{1}$.
The MVM operation in Eq.~\ref{eq:1} originally involves 19 operations and can be reduced to 7 when removing redundant expressions. 
The \gls{DFG} for optimized computation is shown in Fig.~\ref{fig_compilation}e. The red operator and arrow are similar operations to Fig.~\ref{fig_lut} but combined with negative output. After applying loop fission, explicit array access, and \gls{CSE}, the final unrolled code consists of IFM loops, as illustrated in Fig.~\ref{fig_compilation}d.  

To enable vectorization, a sliding window must be stored as a column, achieved by expanding the input channel into a column matrix using the im2col operation (Fig.~\ref{fig_mechanism}b). 
Thus, each iteration of loops in Fig.~\ref{fig_compilation}d generates $C_{out}$ partial dot-products and can be vectorized based on the CAM size.
Finally, the flow annotates the minimum required bitwidth for each level of the DFG. As AP supports arbitrary integer types, we leverage narrow-precision data types for faster addition and reduced energy consumption without affecting correctness.

\subsection{Input mapping and DFG scheduling}
\label{subsec:sched}
Using arrows, Fig.~\ref{fig_mechanism}c shows the dimensions of a typical input mapped to a CAM and its multi-bit RTM cells shown in Fig.~\ref{fig_mechanism}d.   
$F_{h}*F_{w}$ is distributed along the CAM columns, while $H_{out}*W_{out}$ is distributed along the CAM rows. 
Multiple APs can be used to meet the requirements of each layer so that all points in the $H_{out}*W_{out}$ dimension are computed in parallel.  
Fig.~\ref{fig_mechanism}d shows how $N$ bits of the input are stored in a single RTM cell and $C_{in}$ channels can be stored contiguously, sharing the same nanowire and favoring density.
Given the limited number of bits per cell, multiple CAMs are needed to hold a realistic number of channels, thus adding parallelism. 

Finally, instructions are scheduled in two phases: the \textit{channel-wise} DFG, where $C_{in}$ partial OFMs are generated, and \textit{accumulation phase}, where $C_{in}$ partial results are accumulated to form the output channel.
In the first step, the operands in AP instructions (i.e., CAM columns) are treated as registers, similar to the register allocation problem, and solved through graph coloring.
In the second step, we first accumulate partial OFMs locally in the same APs, and we then move partial results from one AP to another in a regular adder tree fashion to favor the parallelism of multiple CAMs. 
The last step also fuses the activation function and stores the OFMs to facilitate the computation of the next layer.

\subsection{Lookup table generation}

We have two options for adder/subtractor implementation: a) in-place, requiring eight cycles, and b) out-of-place, requiring ten cycles. 
Given that the number of cycles/passes needed for in-place operations is smaller than for out-of-place, we maximize the use of in-place operations.
This is done by exploiting the liveliness of a variable in a DFG or by making copies of the result so that in-place operations can be used in the future.
Fig.~\ref{fig_compilation}e shows an example where the operands $x_{6},x_{7}$ and $x_{8}$ are used more than once. 
Thus, their defining operations must store their results in as many columns as they are used by writing in the selected columns in parallel, allowing the subsequent operations to be in place.
Additionally, LUTs for generating negative output are provided at the same cost as the standard ones.

\section{Experimental setup and evaluation results}

As baseline RTM TCAM, we use the \SI{45}{\nano\meter} 256$\times$256 design in~\cite{gnawali2018low}.
It features a search delay under \SI{200}{\pico\s} and per-bit search energy of around \SI{3}{\femto\joule}. We assume 64 domains per nanowire based on \cite{blasing2020magnetic}.
We built a functional simulator that models the architecture presented in Fig.~\ref{fig_mechanism} and estimates performance and energy consumption based on the figure of merits of this CAM design. The functional simulation generates outputs that are analyzed to assess accuracy. We consider a conservative \SI{1}{\pico\joule/\bit} energy consumption for internal data movement at the tile, bank, and global level~\cite{peng2019dnn+}.

\noindent\textbf{Benchmarks:} We evaluate the ImageNet and CIFAR10 datasets on ResNet18 and VGG-9/VGG-11 models, respectively, which were trained with BIPROP~\cite{diffenderfer2021multi} for 100 epochs to identify TWNs that achieve accuracy comparable to a dense full-precision network. We use learned step quantization (LSQ)~\cite{esser2019learned} for quantizing activations to 8 and 4 bits.
~\\\noindent\textbf{Evaluated configurations:} As a comparison baseline, we use an RRAM-based CIM design simulated on DNN+NeuroSim~\cite{peng2019dnn+}, comprising 8-bit weights, 256$\times$256 arrays and 5-bit ADCs. To quantify the individual effect of the proposed transformations on RTM-AP, we evaluate two configurations: \textit{unroll}, implementing loop unrolling, constant weight folding, and custom integer types optimizations, and \textit{unroll+CSE}, including all optimizations from Fig.~\ref{fig_compilation}a.

\subsection{Results summary and comparison to state-of-the-art}
\label{sec:results-summary}
Table \ref{performance-table} summarizes the impact of the sparsity and activation precision on the accuracy, energy consumption, and latency per inference of our proposed AP compared to DNN+NeuroSim~\cite{peng2019dnn+} (crossbar-based) and DeepCAM~\cite{nguyen2023deepcam} (CAM-based). Notably, the 4-bit activation maintains software accuracy while achieving optimal inference performance and energy efficiency. 
Compared to \cite{peng2019dnn+}, our \textit{unroll+CSE} reduces the per inference latency and energy consumption by up to ($\sim$3$\times$, 2.5$\times$) and (1.4$\times$, 1.8$\times$) for ResNet18 and VGG-9, respectively. 
The improvements in performance and energy are partially due to the reduced search/write latency and energy consumption of the RTM-CAMs but are largely attributed to our compiler transformations. For instance, the CSE optimization alone reduces the number of additions by an average of 31\%.
Despite outperforming RTM-AP, DeepCAM faces scalability issues, rendering this comparison unfair. 
DeepCAM depends on large arrays, up to 512 rows and 1024 columns, and the energy efficiency of deeper networks like ResNet18 does not scale as effectively as with LeNet and small VGGs. Also, the accuracy of complex tasks, like ImageNet, is more sensitive to approximation. 
A thorough comparison requires further investigation into the necessary peripheral circuits required by DeepCAM to convert the timing of the match line discharge into digital Hamming distance values, which is beyond the scope of this work. 

\begin{table*}[h]
\centering
\caption{\label{performance-table} Accuracy, energy consumption, latency, required memory and number of operations comparison.}
\begin{tabular}{l r ccc cc rr r rr }
\hline
Network / Dataset       & Sparsity & \multicolumn{3}{l}{Top-1 Accuracy (\%)} & \multicolumn{2}{l}{Energy / Inference ($\mu$J)} & \multicolumn{2}{l}{Latency (ms)}  &  \# Arrays  & \multicolumn{2}{c}{\#Adds/Subs ($1e^{3}$)}  \\
                               &              Actv. prec.             & FP                      & 4-bit   & 8-bit  & 4-bit                     & 8-bit                     & 4-bit           & 8-bit     &  $256\times256$    &  \textit{unroll} & \textit{unroll+CSE}  \\
\hline
ResNet18/ImageNet            & 0.8                       & 70.5                    &   70.6  &  70.6 & 55.04                 & 78.56                 & 2.46             & 4.10      & 49 &   1499   &  931 \\
\hline
ResNet18/ImageNet \cite{peng2019dnn+}              & -           & 70.5                    &   70.0  &  70.0 &  104.92                & 199.90                 & 9.56            & 12.2      & 41 &   n/a   &  n/a \\
    
  \hline
\multirow{2}{*}{VGG-9/CIFAR10}  & 0.85                      & \multirow{2}{*}{93.2}   &   93.5  & 93.5   & 22.80                  & 30.34                 & 1.24             & 2.14   & 4 &   696   & 542 \\
                                & 0.9                       &                         &    93.0 &  93.0  & 16.13                 & 22.11                 & 0.71              & 1.25    &  4   &  470  &  402\\
\hline
VGG-9/CIFAR10 \cite{peng2019dnn+}  & -  & 93.2   &   90.2  & 89.7   & 19.55  &  41.37 & 1.06   & 1.18   & 17 &  n/a   & n/a \\
\hline
\multirow{2}{*}{VGG-11/CIFAR10} & 0.85                      & \multirow{2}{*}{93.6}   &   93.6  &  93.6  & 24.83                 & 36.62                 & 2.47             & 4.24     & 4 &  1390  & 1069  \\
                                & 0.9                       &                         &  93.5 &  93.5  & 18.35                 & 26.86                 & 1.41             & 1.94   & 4  &   929 & 797 \\
\hline
VGG-11/CIFAR10 \cite{nguyen2023deepcam} & n/a              & 93.6   &   90.0  &  90.0  & \multicolumn{2}{c}{0.49}  &       \multicolumn{2}{c}{0.87}    & 24 &  n/a  & n/a  \\
                                
 \hline
\end{tabular}
\vspace{-0.2cm}
\end{table*}

\subsection{Impact on performance and energy consumption}
\label{results}
This section provides a detailed layer-by-layer comparison of the \textit{unroll} and \textit{unroll+CSE} configurations to the baseline DNN+NeuroSim for a more fine-grained analysis. We only focus on ResNet18 (the largest model) for space reasons.
\begin{figure*}
  \centering
  \begin{subfigure}{\textwidth}
    \centering
    \includegraphics[width=0.89\linewidth]{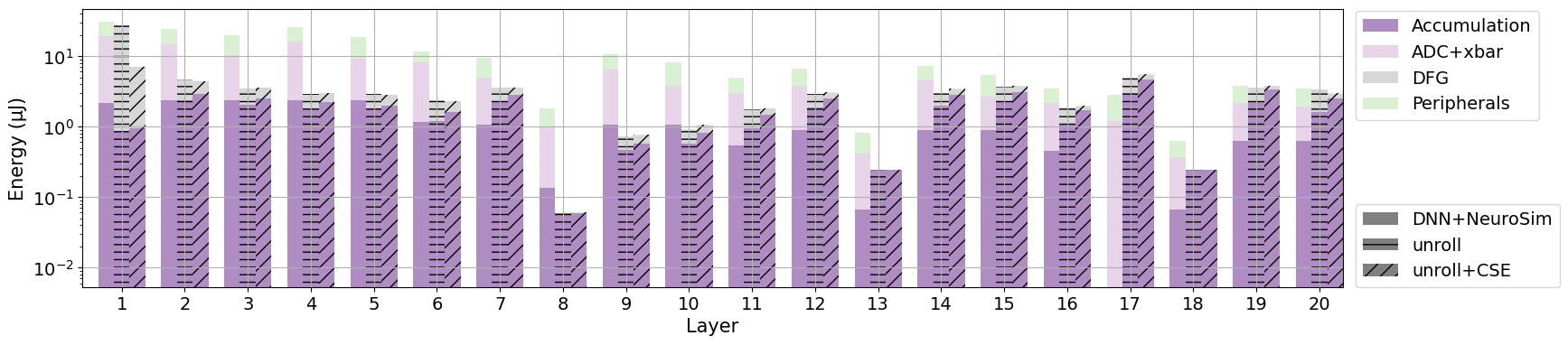}
    \label{fig:subfig1}
  \end{subfigure}%
  \vspace{0.0cm}
  \begin{subfigure}{\textwidth}
    \centering
    \includegraphics[width=0.89\linewidth]{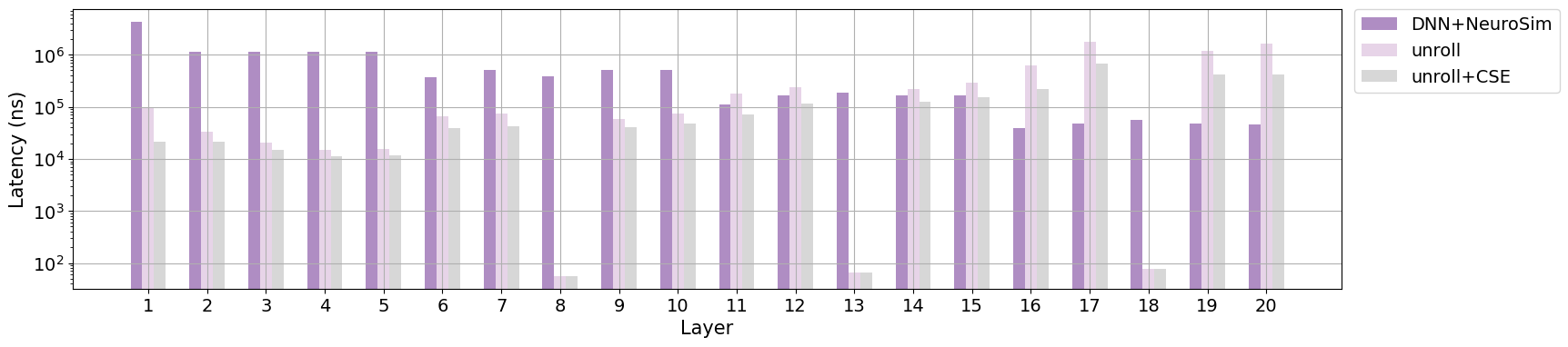}
    \label{fig:subfig2}
  \end{subfigure}
   \vspace{-0.7cm}
  \caption{Layer-by-layer comparison with DNN+NeuroSim~\cite{peng2019dnn+}}
  \label{fig:performance}
  \vspace{-0.5cm}
\end{figure*}
Fig.~\ref{fig:performance} presents the energy consumption and latency of the three configurations, considering the contributions of the individual system components.  
Our solution (\textit{unroll+CSE}) effectively reduces the number of additions (from 1,499K to 931K), thus significantly improving the total energy consumption and latency per image. The reductions are more significant in the first layer because larger kernels (i.e., 7$\times$7) allow for more subexpression elimination.
Also, one can notice that deeper into the network, latency tends to increase due to reduced CAM row utilization as $H_{out}*W_{out}$ decreases. Therefore, the layers 16-20 are slower on RTM-AP compared to \cite{peng2019dnn+}, but still more energy-efficient. 
This could be alleviated, however, by processing multiple images per layer.

Compared to the crossbar-based baseline, \textit{unroll+CSE} can significantly improve end-to-end energy consumption and latency while retaining accuracy. It is worth mentioning that the DNN+NeuroSim model of ResNet-18 does not count with a precise model for weights, activations and ADC quantization, while our modeling requires only activation quantization.  
In DNN+NeuroSim\cite{peng2019dnn+}, \textit{peripherals} account for buffers, digital logic modules (e.g., decoder, switch matrix, mux), and interconnect, while \textit{accumulation} accounts for shift and adders and accumulation units at subarray, tile and global levels. Since our design relies on additional accumulation units, the \textit{accumulation} label denotes the AP energy for computing the \textit{accumulation phase}, while the energy spent in the channel-wise DFG phase is indicated as \textit{DFG}.

\subsection{Impact on data movement and write endurance}
Reducing data movement of partial sums and activations is challenging in current CIM accelerators. We solve this problem by storing activations in CAMs and computing them exclusively with AP operations. Data movement of partial results is not eliminated but it accounts for only 3\% of the total energy consumption. Most of the computation is done locally in each AP, and communication is required only during the \textit{accumulation phase}. In contrast, in crossbar-based design, communication accounts for 41\% of the total energy~\cite{peng2019dnn+}. 

As for write endurance, RTM~\cite{blasing2020magnetic} offers $10^{16}$ write cycles (best compared to other NVMs). In the worst-case scenario, at most two columns are written only once for each in-place or out-of-place operation (Fig. \ref{fig_lut}), which takes 0.8 or 1 ns, respectively. Given that the execution flow is distributed across 256 columns, it is reasonable to rewrite the same column only after approximately 128 operations. This implies writing to the same location roughly every 100 ns, on average, hence resulting in an estimated lifespan of $\sim$31 years.

\section{Conclusion}
We introduced a full-stack solution for efficient NN processing in CAMs. On the software side, our approach leverages compiler transformations applicable to TWNs, ensuring lower arithmetic intensity without compromising accuracy in various large networks. On the hardware front, we enhance existing AP models by integrating RTM cells, harnessing multi-level cell capability and sequential access for bit-serial word-parallel convolution with reduced data transfers within the accelerator. 
This combined approach achieves a remarkable 7.5$\times$ energy efficiency improvement over crossbar-based accelerators.

\section*{Acknowledgments}
This work is funded by the German Research
Council (DFG) through the CO4RTM project (450944241)
and the German Federal Ministry of Education and Research (BMBF, 01IS18026A-D) by funding the competence center for Big Data and AI ScaDS.AI Dresden/Leipzig.
\bibliographystyle{IEEEtran}
\renewcommand{\bibfont}{\small}
\bibliography{main}
\end{document}